# Anisotropic magnetism and band evolution induced by ferromagnetic phase transition in titanium-based kagome ferromagnet SmTi₃Bi₄

Zhe Zheng[1,2†], Long Chen[1,2†], Xuecong Ji[1,2,3†], Ying Zhou[1,2], Gexing Qu[1,2], Mingzhe Hu[1,2], Yaobo Huang[4], Hongming Weng[1,2,5,6*], Tian Qian[1,2,5,6*], and Gang Wang[1,2,5*]

[1] Beijing National Laboratory for Condensed Matter Physics, Institute of Physics, Chinese Academy of Sciences, Beijing 100190, China;
[2] University of Chinese Academy of Sciences, Beijing 100049, China;
[3] School of Physics and Engineering, Henan University of Science and Technology, Luoyang, Henan 471023, China;
[4] Shanghai Synchrotron Radiation Facility, Shanghai Advanced Research Institute, Chinese Academy of Sciences, Shanghai 201210, China;
[5] Songshan Lake Materials Laboratory, Dongguan, Guangdong 523808, China;
[6] CAS Center for Excellence in Topological Quantum Computation, Chinese Academy of Sciences, Beijing 100190, China

Kagome magnets with diverse topological quantum responses are crucial for next-generation topological engineering. The anisotropic magnetism and band evolution induced by ferromagnetic phase transition (FMPT) is reported in a newly discovered titanium-based kagome ferromagnet $SmTi_3Bi_4$, which features a distorted $Ti$ kagome lattice and $Sm$ atomic zig-zag chains. Temperature-dependent resistivity, heat capacity, and magnetic susceptibility reveal a ferromagnetic ordering temperature $T_c$ of 23.2 K. A large magnetic anisotropy, observed by applying the magnetic field along three crystallographic axes, identifies the $b$ axis as the easy axis. Angle-resolved photoemission spectroscopy with first-principles calculations unveils the characteristic kagome motif, including the Dirac point at the Fermi level and multiple van Hove singularities. Notably, a band splitting and gap closing attributed to FMPT is observed, originating from the exchange coupling between $Sm$ $4f$ local moments and itinerant electrons of the kagome $Ti$ atoms, as well as the time-reversal symmetry breaking induced by the long-range ferromagnetic order. Considering the large in-plane magnetization and the evolution of electronic structure under the influence of ferromagnetic ordering, such materials promise to be a new platform for exploring the intricate electronic properties and magnetic phases based on the kagome lattice.

kagome lattice, anisotropic magnetism, band splitting, ferromagnetic phase transition, ARPES

## 1 Introduction

Due to the presence of strong geometric frustration and nontrivial band topology [1, 2], kagome lattice has been acting as a fertile platform to investigate the quantum interactions between geometry, topology, spin, and correlation [3], Nonmagnetic kagome materials have garnered significant attention owing to their potential to exhibit superconductivity [4-8], charge density wave [9-11], nontrivial topological states [12], possible fractionalization [13], and Fermi surface instabilities [14]. With the combination of magnetism, kagome magnets with magnetic ordering would naturally support the tunability of their exotic properties through magnetism, which would result in abundant topological phases and quantum responses [15, 16]. Over the past decade, rich emergent quantum phenomena, including anomalous Hall effects [16-19], anomalous Nernst effects [20-23], negative magnetoresistance from chiral anomaly [24], nontrivial topological states [25-28], charge density waves [29, 30] and so on, have emerged in kagome magnets. Owing to the potential for tuning magnetism through the application of magnetic fields or other controllable parameters, topological kagome magnets hold the promise of hosting highly tunable topological states [15, 16, 27]. This tunability is of paramount importance for the development of next-generation topological engineering.

*Corresponding authors (Hongming Weng, email: hmweng@iphy.ac.cn; Tian Qian, email: tqian@iphy.ac.cn; Gang Wang, email: gangwang@iphy.ac.cn)
† These authors contributed equally to this work.

By introducing various rare-earth elements between stacked kagome lattice, a diverse range of magnetic orderings along with different quantum transport behaviors have been observed in kagome magnets, typically exemplified as $REMn_6Sn_6$ [16, 31, 32], $REV_6Sn_6$ [33, 34], $REV_3Sb_4$ [38] and recently discovered $RETi_3Bi_4$ [39], where $RE$ is a rare-earth element. The coexistence of magnetism stemming from the $Mn$ kagome lattice and the $RE$ atoms results in a spectrum of magnetic ground states for $REMn_6Sn_6$, ranging from distinct ferrimagnetic states ($RE = Gd$-$Ho$) [32, 40] to antiferromagnetic (AFM) order ($RE = Er$, $Tm$, and $Lu$) [32]. In contrast, the $V/Ti$-kagome lattices appear to be nonmagnetic, allowing the $RE$ atoms to dominate the magnetic behavior and give rise to various magnetic states, including nonmagnetic/paramagnetic states in $Er/TmV_6Sn_6$ [33], $YbTi_3Bi_4$ [39], and $YbV_3Sb_4$ [38], as well as AFM states in $REV_6Sn_6$ ($Gd$-$Ho$) [33, 34] and $CeTi_3Bi_4$ [41]. Additionally, there are indications of possible short-ranged order in $PrTi_3Bi_4$ and ferromagnetic (FM) behavior in $TbV_6Sn_6$ [42, 43], $EuV_3Sb_4$ [38], and $NdTi_3Bi_4$ [39]. Among these magnetic configurations, the kagome ferromagnets with either in-plane or out-of-plane magnetization have attracted particular attention due to their potential of hosting Chern gaps or tunable topological phases. Several ARPES results on the electronic structures of $RETi_3Bi_4$ compounds with diverse magnetic properties [39] has been reported recently, distinctive features arising from the kagome sublattice, including Dirac points, multiple van Hove singularities, and flat bands, have been discerned in $NdTi_3Bi_4$, $YbTi_3Bi_4$, and $EuTi_3Bi_4$ [44-46]. Additionally, topological surface states connecting different saddle bands were observed in $EuTi_3Bi_4$ [47]. However, the study of kagome ferromagnets remains rather constrained, underscoring the critical importance of identifying novel topological kagome ferromagnets and explore their distinctive properties and characteristics.

In this work, we present a comprehensive investigation involving single crystal growth, characterization of anisotropic magnetism, and electronic structure analysis in a novel kagome ferromagnet, $SmTi_3Bi_4$. Similar to $REV_3Sb_4$ and $RETi_3Bi_4$, $SmTi_3Bi_4$ encompasses a distorted and nonmagnetic $Ti$-kagome lattice and quasi-one dimensional $Sm$ atomic zig-zag chain. The predominant magnetism, attributed to $Sm$, suggests the presence of a ferromagnetic (FM) order with a critical temperature ($T_c$) of 23.2 K, as indicated by measurements of temperature-dependent resistivity, heat capacity, and magnetic susceptibility. By varying the direction of the applied magnetic field, we observe significant out-of-plane and in-plane anisotropy in the magnetism, with the $b$ axis identified as the easy axis. Through the combination of angle-resolved photoemission spectroscopy (ARPES) and first-principles calculations, we have identified the band characteristics of the kagome lattice, revealing the presence of a Dirac point (DP) at the $\bar{K}$ point at the Fermi level ($E_F$) and multiple van Hove singularities (vHSs) at the $\bar{M}$ point. Furthermore, we have unveiled band splitting induced by the ferromagnetic phase transition (FMPT) via ARPES, originating from the exchange coupling between $Sm$ $4f$ local moments and itinerant electrons of the kagome $Ti$ atoms, as well as the time-reversal symmetry (TRS) breaking [35-37] induced by the long-range ferromagnetic order. The size of the band splitting increases as the temperature decreases. Our results broaden the comprehension of kagome magnets and offer a propitious foundation for delving into the complex electronic and magnetic phases that exist within the kagome lattice framework.

## 2 Materials and methods

**Single Crystal Growth** $SmTi_3Bi_4$ single crystals were grown by a high-temperature solution method using $Bi$ as flux [39]. The as-received $Sm$ ingot (Alfa Aesar, 99.9%) was cut into small pieces, then mixed with $Ti$ powder (99.95%, Alfa Aesar) and $Bi$ granules (99.999%, Sinopharm) with a molar ratio of $Sm : Ti : Bi = 2 : 4 : 12$ in a fritted alumina crucible set (Canfield crucible set) [48] and sealed in a fused-silica ampoule at vacuum. The ampoule was heated to 1073 K over 15 h, held at the temperature for 24 h, and then slowly cooled down to 873 K at a rate of 2 K h$^{-1}$. At 873 K, hexagonal-shaped, shiny-silver single crystals with size up to 5 mm × 5 mm × 0.5 mm were separated from the remaining liquid by centrifuging the ampoule. Considering the possible air sensitivity of the surface, all manipulations and specimen preparation for structure characterization and property measurements were handled in an argon-filled glove box.

**Physical Property Measurements** The temperature-dependent magnetic susceptibility were measured using a vibrating sample magnetometer system (VSM, Quantum Design) under a magnetic field (0.5 T) parallel ($H//bc$) and perpendicular ($H \perp bc$) to the $bc$ plane using both the zero- field-cooling (ZFC) and field-cooling (FC) protocols. Only magnetic susceptibility in FC protocol was measured under a larger field (1 T), and the field-dependent magnetization curves were measured under the magnetic field up to 7 T parallel and perpendicular to the $bc$ plane. The resistivity and heat capacity measurements were carried out using a phys- ical property measurement system (Quantum Design, 9 T). The resistivity was measured using the standard four-probe configuration with the applied current (about 2 mA) parallel to the $bc$ plane. Heat capacity measurement was carried out at temperature ranging from 2.2 K to 200 K at high vacuum (0.01 µbar). To protect the sample from air and moisture, thin film of N-type grease ($\sim 0.05$ mg) was spread to cover the sample in an argon-filled glove box, then the sample was mounted on the square plate of specialized heat capacity puck in air.

**First-principles Calculations** The first-principles calculations are based on density functional theory (DFT) [49] using the Vienna Ab-initio Simulation Package (VASP) [50]. The wave function is expressed with the plane-wave basis set and the exchange and correlation effect are described by the generalized gradient approximation (GGA) with the Perdew-Burke-Ernzerhof (PBE) functional [51, 52]. Projected Augmented Wave - Perdew-Burke-Ernzerhof (PAW-PBE) type of pseudopotential is adopted for $Sm$. The kinetic energy cutoff for the plane-wave basis is set to 500 eV. In the self-consistent calculation, the $\Gamma$-center k-mesh of $7 \times 7 \times 7$ are used for the Brillouin zone integration. Spin-orbit coupling (SOC) is taken into account in all calculations. The parameter Hubbard-U is selected to be 6 eV for localized $4f$ orbitals of $Sm$ atoms, 0 eV for $Ti$. The tight-binding model of $SmTi_3Bi_4$ is constructed by the Wannier90 [53] with $Ti\ 3d$ orbitals, and $Bi\ 6p$ orbitals based on the maximally-localized Wannier functions (MLWF). The surface states and the corresponding Fermi surface are calculated by using the Wannier Tools software package [54].

**ARPES Measurements** ARPES measurements were performed at the "Dreamline" beamline of the Shanghai Synchrotron Radiation Facility (SSRF) with a Scienta Omicron DA30L analyzer and 7eV laser ARPES with a Scienta Omicron DA30L analyzer at the Institute of physics, CAS. The energy and angular resolutions were set to 53 meV and 0.1° at the "Dreamline" and 1 meV and 0.1° at the IOP, CAS. All the samples were cleaved in situ at low temperature and mea-sured in ultrahigh vacuum with a base pressure better than 5 x $10^{-11}$ mbar.

# 3 Results and discussion

$SmTi_3Bi_4$ crystallizes in an orthogonal space group $Fmmm$ (No.69) with lattice parameters $a = 24.900(4)$ Å, $b = 10.3371(16)$ Å, $c = 5.8863(9)$ Å, and $\alpha = \beta = \gamma = 90°$ (More details can be found in Table A1 - A4 and Figure A1 in Supplemental Materials). This compound exhibits a structural motif akin to that found in $REV_3Sb_4$ ($RE = Yb$, $Eu$) [38] and $RETi_3Bi_4$ ($RE = La - Nd$, $Yb$) [39], characterized by a zig-zag chain of $Sm$ atoms and distorted $Ti$ kagome lattice (Figure 1(a)). Figure 1(b) depicts the temperature-dependent in-plane resistivity of $SmTi_3Bi_4$. The residual-resistance ratio ($RRR = \rho_{300K}/\rho_{25K}$) is calculated to be 4.08, indicating excellent crystalline quality of $SmTi_3Bi_4$ single crystal. With current (2 mA) being applied in the $bc$ plane ($I//bc$), the resistivity monotonically decreases with decreasing of temperature, displaying metallic-like behavior with a noticeable drop around $T_c = 23.2$ K. The resistivity can be fitted using the power law: $\rho = \rho_0 + AT^\alpha$, where the value of power $\alpha$ is dependent on the dominant scattering mechanism. Typically, $\alpha$ takes on the following values: 3/2 for diffusive electron motion arising from strong electron correlation [55], 2 for Fermi liquid with moderate electron-electron scattering [56, 57], 3 for dominant $s$-$d$ scattering or electron-magnon scattering [58, 59], and 5 for electron-phonon coupling [60]. Below $T_c$, the resistivity follows a power law with $\alpha_1 = 5.09(15)$, indicating strong electron-phonon coupling. Above $T_c$, the resistivity exhibits a clear non-Fermi liquid behavior with $\alpha_2 = 1.08(1)$, suggesting a linear-in-temperature ($T$ -linear) resistivity, typically associated with quantum critical physics and unconventional superconductivity [61-63]. In comparison with other compounds $RETi_3Bi_4$ ($RE = Yb$, $Pr$, and $Nd$) [39], the smallest value of power $\alpha$ and the highest transition temperature observed in $SmTi_3Bi_4$ imply that stronger electron correlation corresponds to a higher ordering temperature. Remarkably, when an out-of-plane magnetic field is applied, the resistivity and transition temperature remain nearly unaffected as the magnetic field is increased from 1 T to 5 T, suggesting a possible ferromagnetic order.

In comparison with nonmagnetic $YbTi_3Bi_4$, the temperature-dependent specific heat capacity of $SmTi_3Bi_4$ single crystal exhibits a distinct $\lambda$-shape peak at $T_c = 23.2$ K (Figure 1(c)). At high temperature, the specific heat capacity of $SmTi_3Bi_4$ single crystal surpasses the Dulong–Petit limit ($3NR \sim 200$ J mol$^{-1}$ K$^{-1}$, indicated by black dashed line), which is attributed to the more prominent phonon contribution at elevated temperature of N-type grease used for sample protection. By employing the Debye model to fit the specific heat capacity at low temperatures (2 K - 6 K), we determined the Sommerfeld coefficient γ, which is proportional to the density of states (DOSs) around Fermi level ($E_F$) (γ ∝ $g(E_F)$) [64]. For $S\,mTi_3\,Bi_4$, γ is found to be 22.2(8) mJ $K^{-2}$ per formula for $S\,mTi_3\,Bi_4$, a value quite close to that of $YbTi_3\,Bi_4$ (20.7(7) mJ $K^{-2}$ per formula) [39]. The small Som- merfeld coefficient in $S\,mTi_3\,Bi_4$ implies that it is not a heavy fermion system, consistent with our ARPES experimental results and DFT calculations. The density of states near the Fermi level is primarily contributed by $Ti$ and $Bi$, with $f$ electrons situated far from the Fermi level (see Figure 3(g) and **??**). To isolate the specific heat capacity contributed solely by magnons ($C_m$) and corresponding magnetic entropy ($S_m$), we used the specific heat capacity of non- magnetic $YbTi_3\,Bi_4$, which includes contributions from both electrons and phonons, as a reference (see Figure 1(d)). The saturated magnetic entropy is calculated to be 9.24 J $mol^{-1}$ $K^{-1}$ at 30 K, reaching about 79.9% of the theoretical value for $S = 3/2$ ($Rln(2S + 1)$ = 11.56 J $mol^{-1}$ $K^{-1}$) and only 61.8% of that for $S = 5/2$ ($Rln(2S + 1)$ = 14.95 J $mol^{-1}$ $K^{-1}$). This reduction may suggest a departure from the highest spin state of $S\,m^{3+}$ due to the crystal field effect. Furthermore, upon the application of an out-of-plane magnetic field, the peak slightly shifts to higher temperatures, up to 5 T, providing additional evidence for a ferromagnetic order.

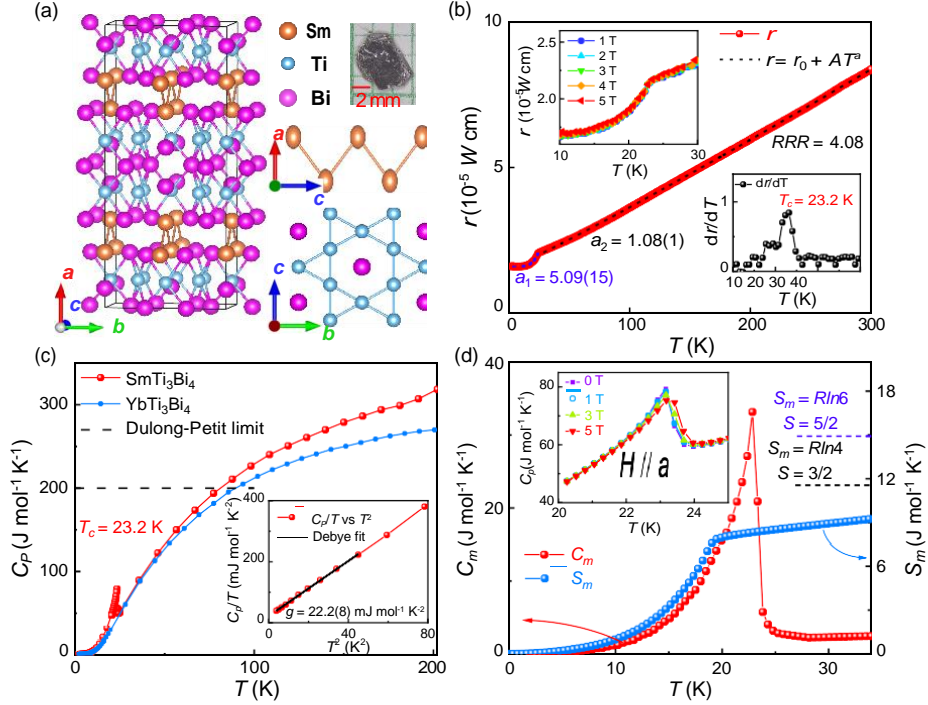

**Figure 1** Crystal structure and phase transition of $S\,mTi_3\,Bi_4$ single crystal. (a) Crystal structure of $S\,mTi_3\,Bi_4$, containing $S\,m$ atomic zig-zag chain viewed along the $b$ axis and $Ti$-kagome lattice in $Ti$-$Bi$ layer viewed along the $a$ axis. Upper right part shows an optical photograph of an as-grown single crystal with a grid size of 2 mm. (b) Temperature-dependent in-plane resistivity of $S\,mTi_3\,Bi_4$ single crystal with current in the $bc$ plane ($I///bc$). The dashed lines show the power law fittings. Upper inset shows the temperature-dependent resistivity under out-of-plane magnetic field (1 T - 5 T). Lower inset shows the first derivative of resistivity ($dp/dT$). (c) Temperature-dependent specific heat capacity of $S\,mTi_3\,Bi_4$. The specific heat capacity of nonmagnetic $YbTi_3\,Bi_4$ is taken as reference. The inset shows $C_P$ vs $T^2$ plot and corresponds fittings (black lines) using Debye model. (d) Specific heat capacity contributed by magnon ($C_m$) and corresponding magnetic entropy ($S_m$). The dashed lines are theoretical saturated magnetic entropy with different spin states ($S = 5/2$ or $3/2$). The inset shows the specific heat capacity of $S\,mTi_3\,Bi_4$ under different out-of-plane magnetic field (1 T - 5 T).

Figure 2(a-c) shows the temperature-dependent magnetic moment of single-crystal $S\,mTi_3\,Bi_4$ when subjected to a magnetic field parallel to the $a$ axis, $c$ axis and $b$ axis, respectively. Under a weak magnetic field (0.1 T), a discernible bifurcation between the zero-fielding-cooling (ZFC) and field-cooling (FC) curves can be clearly resolved at low temperature, with magnetic moment exhibiting a sharp increase around $T_c$ = 23.2$K$, indicative of a potential ferromagnetic (FM) order. Under a large magnetic field (7 T), the magnetic moment demonstrates an augmented response and exhibits a pronounced enhancement around $T_c$ upon cooling. Additionally, it adheres to the Curie-Weiss law χ = $χ_0$ + $C/(T − θ)$ at elevated temperature (black lines). Here, $χ_0$ represents the temperature-independent contribution, encompassing the diamagnetic component of the orbital magnetic moment, $C$ denotes the Curie constant, and θ signifies the Curie temperature. The fitted Curie temperatures for $H///a$ and $H///b$ are notably positive (85 K for $H///a$ and 23 K for $H///b$), showing

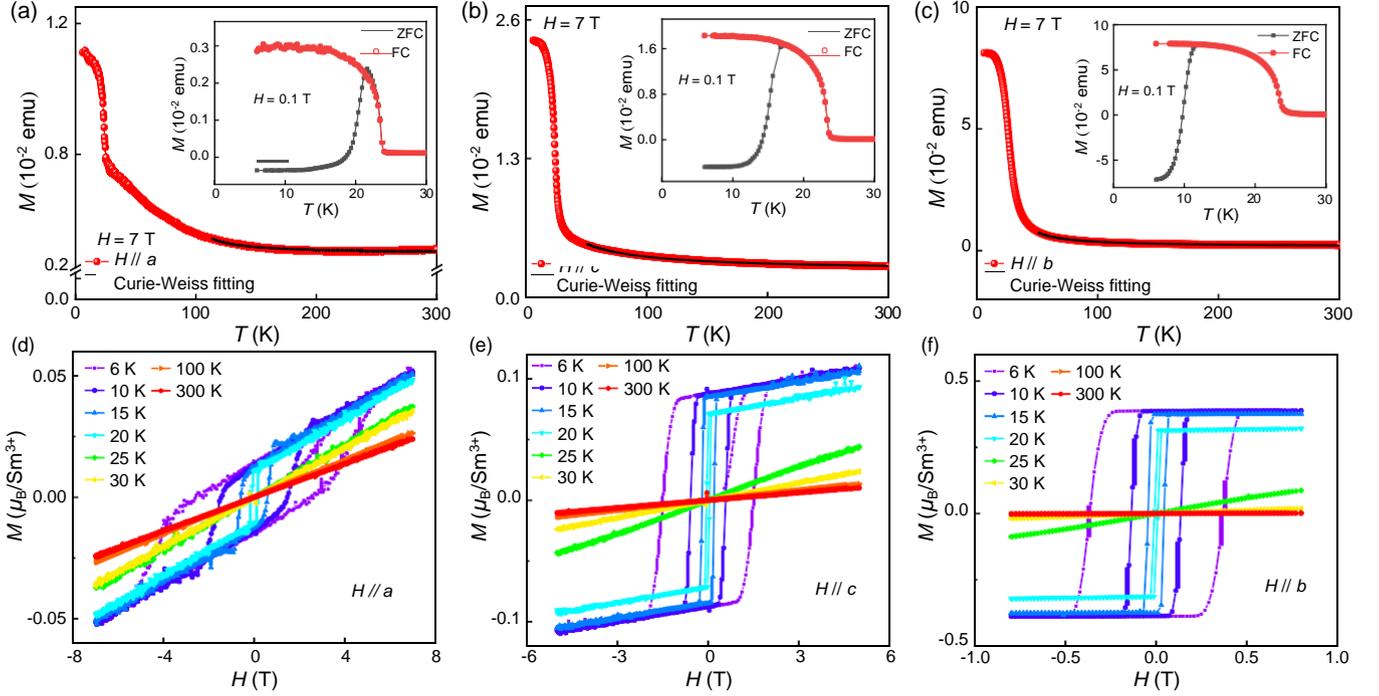

**Figure 2** Ferromagnetic order in *S m*Ti₃Bi₄ single crystal. Temperature-dependent magnetic moment of *S m*Ti₃Bi₄ with a large magnetic field (7 T) parallel to the (a) *a* axis (*H*//*bc*), (b) *c* axis (*H*//*bc*, θ = 269°), and (c) *b* axis (*H*//*bc*, θ = 178°). The black lines are Curie-Weiss fitting curves. The insets show the corresponding ZFC and FC curves under a small magnetic field (0.1 T). Field-dependent magnetization curves at different temperatures of *S m*Ti₃Bi₄ under a magnetic field parallel to the (d) *a* axis, (e) *c* axis, and (f) *b* axis

strong FM interactions. For *H*//*c*, the fitted Curie temperature is close to zero (-3 K), indicating a largely reduced FM interaction along the *c*-axis. The anisotropy FM interaction may arise from the quasi-one-dimensional zig-zag arrangement of *S m* atoms. By employing the equation $\mu_{eff} = \frac{8C}{n}$, where *n* represents the number of magnetic atoms, the effective moments are calculated to be 0.20(3) $\mu_B$ for *H*//*a*, 0.71(1) $\mu_B$ for *H*//*b*, and 0.64(1) $\mu_B$ for *H*//*c*, respectively. Particularly, the effective moment for *H*//*b* is quite close to the theoretical effective moment of *S m*³⁺ (0.84 $\mu_B$). Furthermore, based on the value of magnetic susceptibility under 7 T at 6 K, where $\chi_{H//b}$ (3.2 × 10⁻² emu mol⁻¹ Oe⁻¹) >$\chi_{H//c}$ (0.9 × 10⁻² emu mol⁻¹ Oe⁻¹) >$\chi_{H//a}$ (4.2 × 10⁻³ emu mol⁻¹ Oe⁻¹), it can be concluded that the easy plane is the *bc* plane and the easy axis aligns with the *b* axis. (The spin structure in the schematic crystal structure is shown in Figure A3 in Supplemental Materials.) The FM ordering occurring at *T_c* is unequivocally affirmed by field-dependent magnetization curves, which exhibit conspicuous hysteresis loops at low temperatures and gradually diminish as the temperature rises beyond *T_c* (Figure 2(d-f). Notably, for *H*//*b*, the smaller coercive magnetic field (*H_c, b*// ~ 0.5 T <*H_c, c*//~ 2.5 T <*H_c, a*// ~ 5 T) and the larger saturation moment (μ_sat, b// ~ 0.5 $\mu_B$ >μ_sat, c// ~ 0.1 $\mu_B$ >μ_sat, a// ~ 0.05 $\mu_B$) further confirm the easy axis as the *b* axis. Taking into account the *b* axis as the easy axis and the presence of FM ordering below *T_c*, the *S m* atomic zig-zag chains, with their substantial in-plane magnetization sandwiched between *Ti*-kagome lattices, are expected to induce tunable topological phases. This is because the significant in-plane magnetization would close the gaps in topological kagome magnets. Analogous scenarios, involving long-range magnetic order on low-dimensional lattices nestled within topological lattices, have recently been predicted or observed in other systems. Examples include magnetism-tuned topological phases in the square-net compound *ErAsS* , resulting in exotic hourglass surface states [65], and anomalous Hall conductivity in the potential Chern insulator *TbV₆S n₆* [43].

To investigate the electronic structure of *S m*Ti₃Bi₄, we conducted angle-resolved photoemission spectroscopy (ARPES) measurements in conjunction with first-principles calculations. Figure 3(a) shows the bulk Brillouin zone along with its projection on the (100) surface. Due to a slight distortion of the kagome lattice, the surface Brillouin zone has a quasi-hexagonal shape with two longer sides and four shorter sides. The experimental and calculated Fermi surfaces and band dispersions in the paramagnetic state are presented in Figure 3(b-e). The experimental Fermi surfaces in Figure 3(b) exhibit a quasi-sixfold symmetry overall, except for the outer Fermi surface around the Γ̄ point, which possesses a twofold symmetry. These findings align well with the calculated results shown in Figure 3(c). The breaking of the

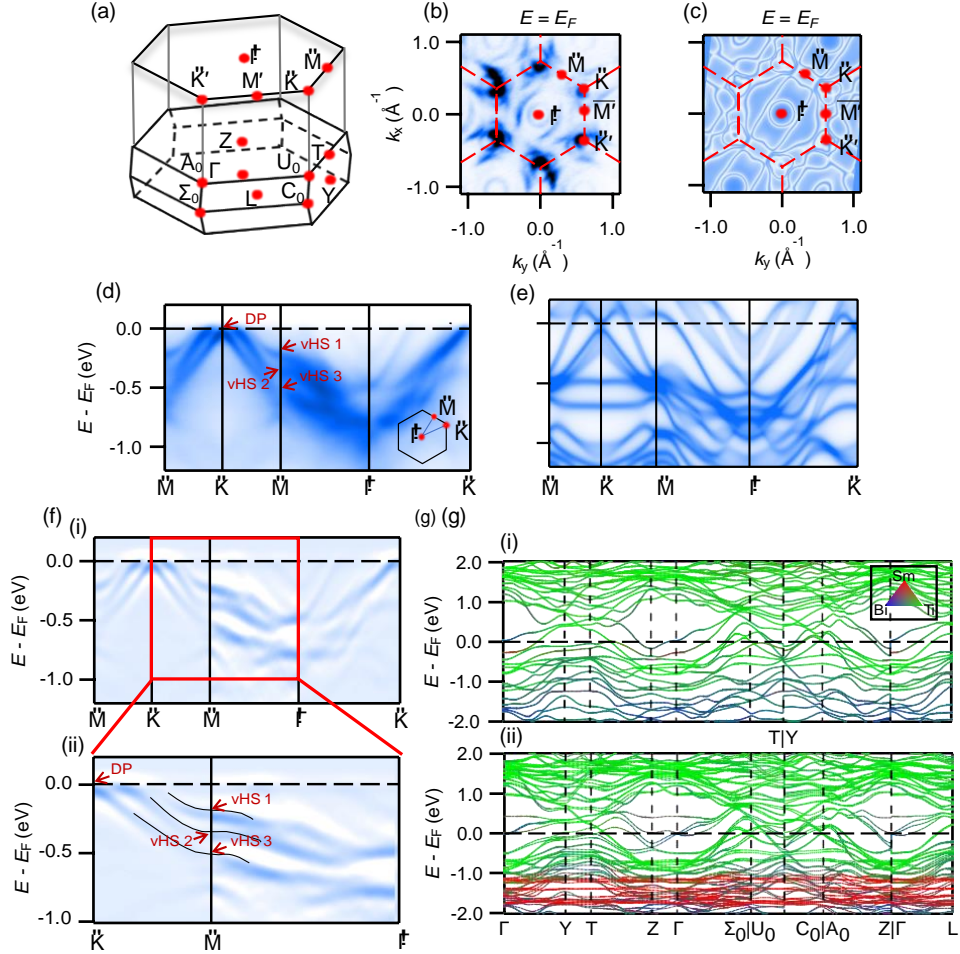

**Figure 3** Calculated and observed Fermi surface and band structure of $SmTi_3Bi_4$. (a) Three-dimensional Brillouin zone of $SmTi_3Bi_4$ and its projection on the (100) surface with high-symmetry points indicated. (b) Constant-energy contours at the Fermi level ($E_F$) measured at 125 eV and 38.8 K. The red dashed quasi-hexagon represents the Brillouin zone. (c) Calculated two-dimensional projected Fermi surface. (d) ARPES intensity plot along the surface high-symmetry $\bar{M}$-$\bar{K}$-$\bar{M}$-$\bar{\Gamma}$-$\bar{K}$ path measured at 125eV and 38.8 K. (e) The calculated surface projection of bands along high-symmetry $\bar{M}$-$\bar{K}$-$\bar{M}$-$\bar{\Gamma}$-$\bar{K}$ direction, which shows the typical bands characteristic resulting from the kagome motif. (f) (i) Second derivative of the ARPES intensity plot in (d). (ii) Zoom-in plot of the red region in (i). Dashed lines are guide to eyes of the band dispersions of multiple van Hove singularities (vHSs). (g) Bulk electronic structure in DFT Calculations. (i) Paramagnetic state, (ii) Ferromagnetic state. Red, green, and blue represent the contributions from $Sm$, $Ti$, and $Bi$ atomic orbitals, respectively. The 4f electron was treated as core electron and local moment in paramagnetic state and ferromagnetic state, respectively.

sixfold rotation symmetry signifies that the insertion of zigzag $Sm$ atomic chains has a noteworthy impact on the low-energy electronic states. The measured band dispersions in Figure 3(d) display clear band characteristics of the kagome lattice, including the presence of a Dirac point at the $\bar{K}$ point, very close to the Fermi level, and multiple van Hove singularities at the $\bar{M}$ point. In the kagome lattice with time-reversal symmetry, the Dirac point at the $K$ point is symmetry protected in the absence of spin-orbit coupling. As the kagome lattice in $SmTi_3Bi_4$ is slightly distorted, the Dirac point opens a small gap. Three saddle-like band structures at the point can be clearly identified in Figure 3(f), i.e., hole-like bands along $\bar{\Gamma} - \bar{M}$ and electron-like bands along $\bar{K} - \bar{M}$, indicating multiple van Hove singularities. Bulk electronic structures in Figure 3(g) demonstrate that the band structure near the

Fermi level is primarily contributed by the $Ti$-kagome lattice. The non-dispersive 4f-electron bands can be identified at 135 eV, and a strong energy resonance occurs. (See Figure A6 in Supplemental Materials.) These experimental observations closely match the calculated results.

To explore the impact of time-reversal symmetry breaking on the electronic structure of $SmTi_3Bi_4$, we conducted temperature-dependent ARPES measurements across $T_c$. Significant changes are detected in the band dispersions as a function of temperature, as shown in Figure 4(a). In the paramagnetic state, we identify two electron-like bands (referred to as α and β) and a hole-like band (referred to as γ) in the vicinity of the $\bar{\Gamma}$ point. As the temperature decreases to 17.9 K, slightly below $T_c$ = 23.2 K, we observe an energy splitting in the α band. In Figure 4(b), the energy dis-tribution curves (EDCs) at the $\bar{\Gamma}$ point reveal that the magnitude of the splitting gradually increases with further decreasing temperature. Furthermore, there is a noticeable gap between the $\beta$ and $\gamma$ bands above $T_c$. As the temperature decreases, this gap gradually narrows and eventually becomes negligible, as depicted in Figure 4(a). The observed gap closing is consistent with the calculated results presented in Figure 4(d). These results indicate that time-reversal symmetry breaking has a profound impact on the low-energy electronic states. In $SmTi_3Bi_4$, the time-reversal symmetry breaking originates from the long-range FM order of $Sm$ $4f$ local moments, while the low-energy electronic states are primarily contributed by the itinerant $3d$ electrons of $Ti$ atoms that form the kagome lattice. The pronounced variations in band dispersions signify strong exchange coupling, leading to significant Zeeman splitting. These findings highlight the intimate connection between magnetism and electronic structure, pro-viding essential insights for a deeper understanding of the physical properties of $SmTi_3Bi_4$.

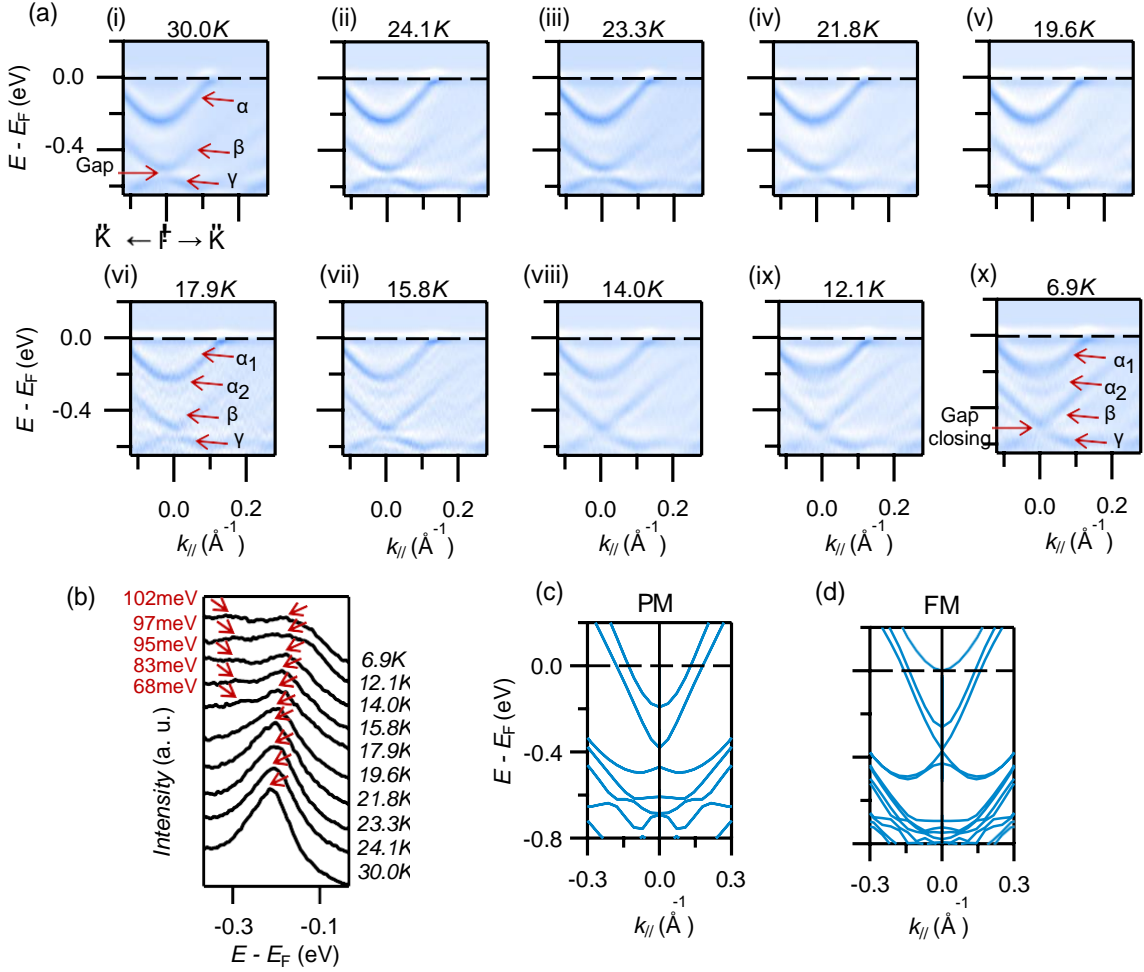

**Figure 4** Temperature dependence of band dispersions around $\bar{\Gamma}$ via laser ARPES measurement. (a) (i)-(x) Second derivative of the ARPES data along $\bar{K}$-$\bar{\Gamma}$-$\bar{K}$ at different temperatures between 30 K and 6.9 K. The red arrows mark the electron bands ($\alpha$ and $\beta$), the hole band ($\gamma$), the energy gap, the gap closing, and the two electron bands ($\alpha_{1,2}$) resulting from the energy splitting, which both emerge as the temperature decreases. (b) Energy distribution curves (EDCs) near the $\bar{\Gamma}$ point of the data in (a). Red arrows are guide to eyes of the peaks of the EDCs. Calculated electronic structures along high-symmetry $\bar{K}$-$\bar{\Gamma}$-$\bar{K}$ line for $SmTi_3Bi_4$ in (c) Paramagnetic state, (d) Ferromagnetic state.

## 4 Conclusions

In summary, we have successfully synthesized a novel $Ti$-based kagome magnet, $SmTi_3Bi_4$, which exhibits characteristic kagome lattice features and significant magnetic anisotropy. Our measurements of temperature-dependent resistivity, heat capacity, and magnetic susceptibility confirm the low-temperature (below 23.2 K) ferromagnetic behavior of this material, characterized by anisotropic magnetism with the $b$ axis as the easy axis of magnetization. Our ARPES measurements and first-principles calculations reveal that the Fermi surfaces are characterized by quasi-sixfold rotation symmetry and twofold symmetry. Distinctive band structures originating from the $Ti$-kagome lattice are clearly observed.

Notable features include the presence of a Dirac point at the $\bar{K}$ point and multiple van Hove singularities at the $\bar{M}$ point. Furthermore, we observed pronounced band splitting induced by FMPT. This suggests strong exchange coupling between local moments and itinerant electrons. The gap closing signifies that long-range ferromagnetic ordering breaks time-reversal symmetry, resulting in significant Zeeman splitting. The study on $SmTi_3Bi_4$ provides a crucial foundation for a deeper understanding of electronic properties and its potential applications.

*Zhe Zheng thanks Shunye Gao, Bei Jiang, Shenggen Cao, Renjie Zhang, Junde Liu and Mojun Pan for their assistance in ARPES experiment and data processing. This work was supported by the National Natural Science Foundation of China (Grant No. 51832010, No. 11888101, No. 11925408, No. 11921004, No. 12188101, and No. U22A6005), the National Key Research and Development Program of China (Grant No. 2022YFA1403900, No. 2018YFE0202600, and No. 2022YFA1403800), the Ministry of Science and Technology of China (Grant No. 2022YFA1403800), the Chinese Academy of Sciences (Grant No. XDB33000000) and the Informatization Plan of the Chinese Academy of Sciences (Grant No. CASWX2021SF-0102), the Strategic Priority Research Program of Chinese Academy of Sciences (Grant Nos. XDB33000000 and XDB28000000), the Synergetic Extreme Condition User Facility (SECUF), and the "Dreamline" beamline of Shanghai Synchrotron Radiation Facility (SSRF). All authors contributed to the scientific planning and discussions.*

# Supplemental Materials

## A1 Structure Characterization and Composition Analysis

X-ray diffraction data were obtained using a PANalytical X'Pert PRO diffractometer ($Cu$ $K_\alpha$ radiation, $\lambda = 1.54178$ Å) operated at 40 kV voltage and 40 mA current with a graphite monochromator in a reflection mode ($2\theta = 5°\text{-}100°$, step size = $0.017°$). Indexing and Rietveld refinement were performed using the DICVOL91 and FULLPROF programs [1]. Single crystal X-ray diffraction (SCXRD) data were collected using a Bruker D8 VENTURE with $Mo$ $K_\alpha$ radiation ($\lambda = 0.71073$ Å) at 280 K for $S$ $mTi_3Bi_4$. The structure was solved using a direct method and refined with the Olex2 [2] and Jana2020 [3] package. The morphology and chemical composition were characterized using a scanning electron microscope (SEM-4800, Hitachi) equipped with an electron microprobe analyzer for semiquantitative elemental analysis in energy-dispersive spectroscopy (EDS) mode. Three spots in different areas were measured on each crystal using EDS.

**Table A1** Crystallographic data and structure refinement of $S$ $mTi_3$ $Bi_4$

| Empirical formula | $S$ $mTi_3$ $Bi_4$ |
|---|---|
| f.u. weight (g mol⁻¹) | 1129.97 |
| Space group / Z | $Fmmm$ (No.69) / 4 |
| $a$ (Å) | 24.900(4) |
| $b$ (Å) | 10.3371(16) |
| $c$ (Å) | 5.8863(9) |
| $\alpha$, $\beta$, $\gamma$ (°) | 90 |
| $V$ (Å³) | 1515.1(4) |
| $d_{cal}$ (g cm³) | 4.954 |
| Refl. Collected / unique | 2997 / 535 |
| $R_{int}$ | 0.0570 |
| Goodness-of-fit | 1.126 |
| $R_1$/$wR_2$ ($I > 2\sigma(I)$) | 0.1025 / 0.3321 |
| $R_1$/$wR_2$ (all) | 0.1068 / 0.3418 |

**Table A2** Quantitative analysis of chemical composition for $S$ $mTi_3$ $Bi_4$ single crystals. For each sample, the molar ratio of composed elements is measured at three different locations. The final stoichiometry of $S$ $mTi_3$ $Bi_4$ is determined to be $S$ $m$ : $Ti$ : $Bi$ = 0.95(3) : 3.00(6) : 4.16(6) by the statistical average and standard deviation (SD) of the overall EDS data.

| Atom | $S$ $m$ (mol%) | $Ti$ (mol%) | $Bi$ (mol%) |
|---|---|---|---|
| SA#-1 | 11.66 | 35.30 | 53.04 |
| SA#-2 | 11.79 | 36.61 | 51.60 |
| SA#-3 | 11.79 | 37.30 | 51.59 |
| SB#-1 | 11.74 | 37.33 | 50.93 |
| SB#-2 | 11.18 | 38.02 | 50.80 |
| SB#-3 | 11.90 | 36.79 | 51.31 |
| SC#-1 | 12.20 | 37.79 | 50.41 |
| SC#-2 | 11.95 | 36.80 | 51.62 |
| SC#-3 | 11.95 | 37.59 | 50.46 |
| Average | 11.68 | 37.01 | 51.31 |
| SD | 0.35 | 0.78 | 0.80 |

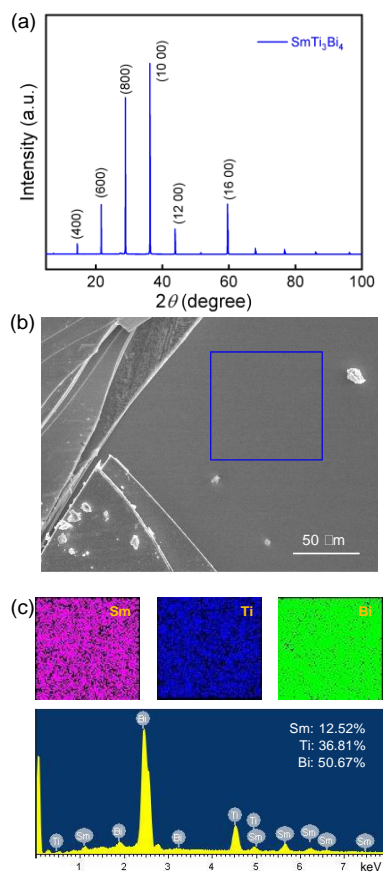

**Figure A1** Crystal structure and chemical composition. (a) Single crystal XRD pattern, showing ($h00$) ($h$ is an even number) diffraction peaks. (b) SEM image of $S$ $mTi_3$ $Bi_4$ at a scale of 100 $\mu m$. (c) Elemental mapping and elemental analysis by EDS, showing a homogeneous distribution of $Sm$, $Ti$, and $Bi$.

**Table A3** Atomic coordinates and equivalent isotropic displacement parameters for $S\,mTi_3\,Bi_4$

| Atom | Wyck. | S ym. | x/a | y/b | z/c | Occ. | U(eq)(Å²) |
|------|-------|-------|------|------|------|------|-----------|
| $S\,m$ | 8g | 2mm | 0.6958(2) | 0.5000 | 0 | 1.0 | 0.0101(11) |
| Ti1 | 8g | m | 0.5933(4) | 0.5000 | 0.5000 | 1.0 | 0.0060(20) |
| Ti2 | 16l | m | 0.5961(3) | 0.2500 | 0.2500 | 1.0 | 0.0052(17) |
| Bi1 | 8h | m | 0.5000 | 0.3306(2) | 0.5000 | 1.0 | 0.0061(9) |
| Bi2 | 16o | m | 0.6889(1) | 0.3399(2) | 0.5000 | 1.0 | 0.0068(9) |
| Bi3 | 8g | m | 0.5691(1) | 0.5000 | 0 | 1.0 | 0.0063(9) |

**Table A4** Anisotropic displacement parameters for $S\,mTi_3\,Bi_4$. The anisotropic displacement factor exponent takes the form: $-2\pi^2(h^2a^2U_{11} + \ldots + 2hkabU_{12})$

| Atom | $U_{11}$ (Å²) | $U_{22}$ (Å²) | $U_{33}$ (Å²) | $U_{23}$ (Å²) | $U_{13}$ (Å²) | $U_{12}$ (Å²) |
|------|------|------|------|------|------|------|
| $S\,m$ | 0.0053(17) | 0.0034(18) | 0.0220(30) | 0 | 0 | 0 |
| Ti1 | 0.0240(60) | 0.0160(50) | -0.0220(50) | 0 | 0 | 0 |
| Ti2 | 0.0090(40) | 0.0060(40) | 0.0010(40) | 0 | 0 | 0.0010(30) |
| Bi1 | 0.0083(15) | 0.0043(14) | 0.0055(17) | 0 | 0 | 0 |
| Bi2 | 0.0047(13) | 0.0029(13) | 0.0127(17) | 0.0008(5) | 0 | 0 |
| Bi3 | 0.0034(13) | 0.0029(13) | 0.0126(18) | 0 | 0 | 0 |

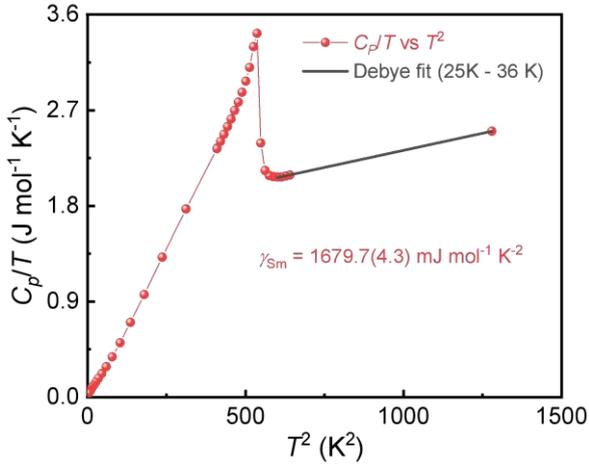

**Figure A2** Debye fitting beyond $T_C$. $C_P/T$ vs $T^2$ plot and corresponding fitting (black lines) using Debye model at high temperature (25 K - 36 K).

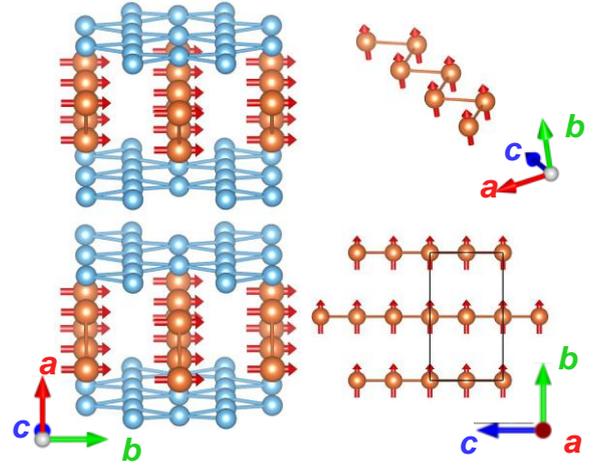

**Figure A3** Schematic crystal structure with ferromagnetic ordering in $S\,mTi_3\,Bi_4$. Only $S\,m$ atoms in the zig-zag chain and $Ti$-kagome lattice are shown for clarity.

## A2 The constant-energy contours and band structure of $S\,mTi_3Bi_4$

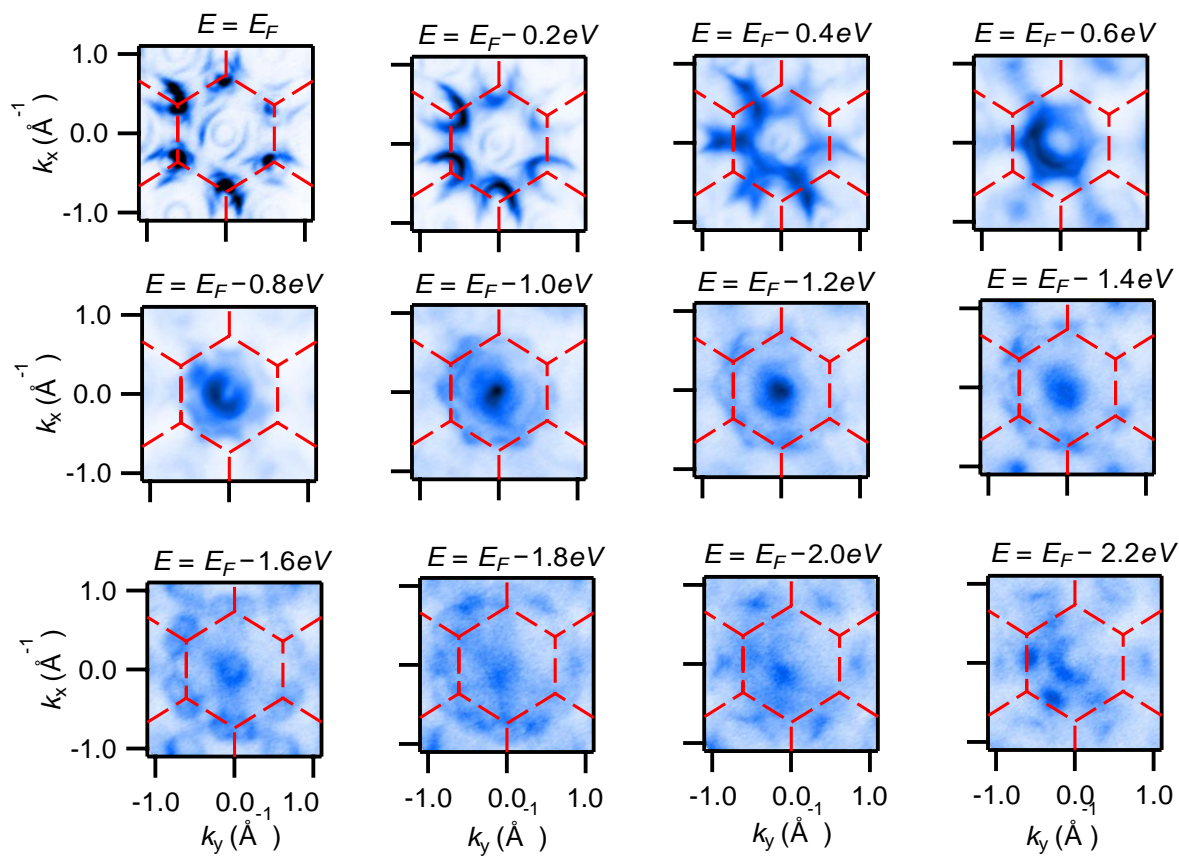

**Figure A4** The constant-energy contours of $SmTi_3Bi_4$. The constant-energy contours were observed at different binding energies, measured at 125 eV and 38.8 K. The features of two-fold symmetry and quasi-sixfold rotation symmetry were detected. The red dashed quasi-hexagon represents the (100) surface projection of the Brillouin zone.

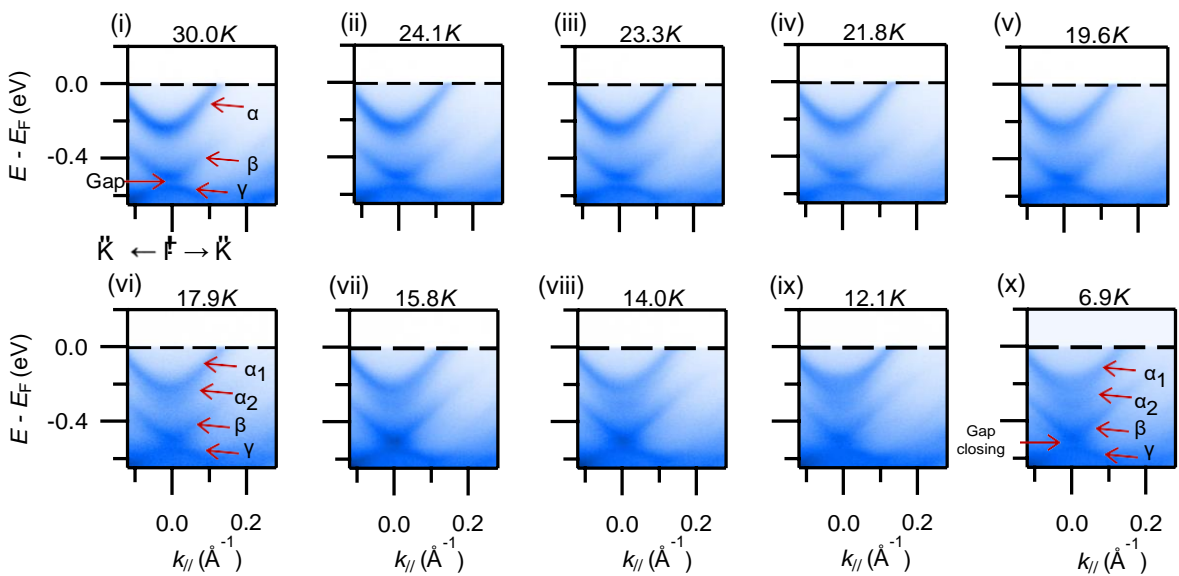

**Figure A5** Temperature dependence of band dispersions around $\bar{\Gamma}$ via laser ARPES measurement. (i)-(x) Intensity plot of temperature-dependent ARPES data along $\bar{K}$-$\bar{\Gamma}$-$\bar{K}$ at different temperatures between 30 K and 6.9 K ranging from 30 K to 6.9 K. The red arrows mark the electron bands ($\alpha$ and $\beta$), the hole band ($\gamma$), the energy gap, the gap closing, and the two electron bands ($\alpha_{1,2}$) resulting from the energy splitting, which both emerge as the temperature decreases.

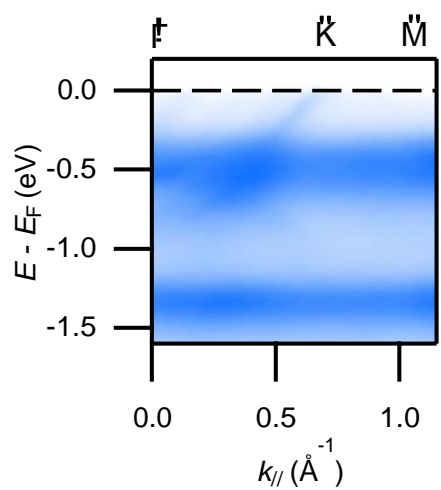

**Figure A6** Intensity plot of the resonance ARPES data along $\bar{\Gamma}$-$\bar{K}$-$\bar{M}$ measured at 135 eV and 41.6 K, showing the $4f$ states around -0.5 eV and -1.4 eV.